\documentclass[letterpaper,12pt]{article}
\usepackage{amssymb,epsfig,setspace} 
\begin{document}

\hyphenation{ge-ne-ra-tes}
\hyphenation{me-di-um  as-su-ming pri-mi-ti-ve pe-ri-o-di-ci-ty}
\hyphenation{e-q-ua-ti-on Er-go-dic}
\hyphenation{wa-ves di-men-si-o-nal ge-ne-ral the-o-ry sca-t-te-ri-ng}
\hyphenation{di-f-fe-r-ent tra-je-c-to-ries e-le-c-tro-ma-g-ne-tic pho-to-nic}

\title{On unorthodox solutions of the Bloch equations}

\author{
Alexander Moroz\thanks{wavescattering@yahoo.com}
}

\date{
Wave-scattering.com
} 

\maketitle

\begin{center}
{\large\sc abstract}
\end{center}
A systematic, rigorous, and complete investigation of the Bloch equations
in time-harmonic driving classical field is performed.
Our treatment is unique in that it takes full advantage of the
partial fraction decomposition over real number field, which 
makes it possible to find and classify all analytic solutions.
Torrey's analytic solution in the form of 
exponentially damped harmonic oscillations 
[Phys. Rev. {\bf 76}, 1059 (1949)] is found to 
dominate the parameter space, which justifies its use 
at numerous occasions in magnetic resonance and in quantum 
optics of atoms, molecules, and quantum dots. 
The unorthodox solutions of the Bloch equations, which do not
have the form of exponentially damped harmonic oscillations, 
are confined to rather small detunings 
$\delta^2\lesssim (\gamma-\gamma_t)^2/27$ and 
small field strengths $\Omega^2\lesssim 8 (\gamma-\gamma_t)^2/27$, where
$\gamma$ and $\gamma_t$ describe 
decay rates of the excited state (the total
population relaxation rate) and of the coherence, respectively.
The unorthodox solutions being readily accessible experimentally 
are characterized by rather featureless time dependence.

\vspace*{0.08cm}

\noindent Keywords: analytic solutions * Bloch equations *
two-level system * time-harmonic driving field

\vspace*{0.08cm}

\noindent PACS codes: 03.67.-a, 42.50.-p, 82.56.-b, 33.50.-j

\newpage

\section{Introduction}
\label{sc:intr}
The Bloch equations in time-harmonic driving classical field 
have been employed over several decades as 
an important tool in studies of many different physical phenomena 
\cite{Bl,Tr,FVH,HM,MK,McH,WLK,AE,Ya,BMO,BTL,TLB,TJB,BNB,FMR,DS}. 
The equations [see (\ref{bleq}) below] underline the theory
of magnetic resonance \cite{Bl,Tr,FVH,HM,MK} and the quantum optics 
of a two-level atom (molecule, spin, ion, etc) driven 
by a classical field \cite{FVH,McH,WLK,AE,Ya,BMO,BTL,TLB,TJB,BNB}. 
The latter problem  
is one of the most discussed and is at the heart of 
the theory of self-induced transparency
\cite{McH,AE}, the susceptibility of an ensemble of 
atoms (spins, ions, etc) in gain media \cite{Ya}, 
and a number of other optical phenomena \cite{AE,Ya,BMO,BTL,TLB,TJB}. 
Recently the problem has been extensively
studied in connection with the proposed use of 
atoms and molecules as a triggered single-photon emitter \cite{BTL,TJB},
a single-photon emission of resonantly driven 
semiconductor quantum dot in a microcavity \cite{FMR},
 and decoherence in dc SQUID phase qubits \cite{DS}.
Advances in the fabrication of single defect centres in diamond
enable one to investigate two-level diamond-based single-photon emitters
at room temperatures \cite{JW,ACS}.
At the same time the single defect centres enable
nanoscale magnetic sensing, and hence a
nanoscale imaging magnetometry, with an 
individual electronic spin in diamond
under ambient conditions \cite{TCC,ML,BJW}.

The Bloch equations form
a linear system of three ordinary differential equations 
[see (\ref{bleq}) below],
which can be formally solved in the following three steps:
\begin{itemize}

\item 
(i) applying the {\em Laplace transform}, whereby the system of
differential equations reduces to a linear algebra problem; 

\item (ii) solving the ensuing linear algebra problem
(for instance, 
by means of {\em Cramer's rule}); 

\item (iii) applying 
the {\em inverse Laplace transform}.
\end{itemize}
That was also the original Torrey's approach \cite{Tr,WLK,AE},
who established a general time dependence of each Bloch 
variable $u_l$, $l=1,\,2,\,3$, in the form of 
{\em exponentially damped harmonic oscillations}
\begin{equation}
u_l(t) = A_0 + A_1 e^{-\kappa_1 t} 
            + A_2 e^{-bt}\cos(st) + (A_3/s) e^{-bt}\sin(st).
\label{trsl}
\end{equation}
In order to determine explicit analytic solutions,
essential to Torrey's approach was to find out 
a {\em negative real} root $-\kappa_1$ (it always exists
- see section \ref{sec:mt}) 
of a characteristic determinant $\Delta$ 
[see  (\ref{dtp}) below] of the Bloch equations. 
Assuming the knowledge of $\kappa_1$, 
Torrey \cite{Tr} determined the constants $A_0$, $A_1$ as certain limits 
of Laplace transforms [see (\ref{a0f}), (\ref{a1f}) below], 
and determined the remaining constants $A_2$, $A_3$ using the 
initial conditions and the knowledge of $A_0$ and $A_1$ 
(see Eqs. (45)-(47) of \cite{Tr}). 
The root parameters $b$ and $s$ were then determined
on comparing expansion coefficients 
of the cubic polynomial $\Delta(p)$ 
in the Laplace transform variable $p$ against 
corresponding  expressions of the coefficients in terms of cubic roots
 [see Eq. (48) of \cite{Tr} and (\ref{bsd}) below]. 
Not aware of {\em Cardano's formula} \cite{Cf} - 
a jewel of the mathematics of the $16$th century \cite{Cr} -
Torrey managed to formulate explicit analytic solutions
merely in three special situations \cite{Tr}:
\begin{itemize}

\item {\em strong collisions:} $\gamma=\gamma_t$ ({\bf T1});
\item {\em exact resonance:} $\delta=0$ ({\bf T2});
\item {\em intense external field:} $\Omega/\gamma_t \gg 1$ ({\bf T3}).

\end{itemize}
Additionally, it had long remained unnoticed that 
Torrey's general solution (\ref{trsl}) had been confined to 
the parameter range when the discriminant 
$D$ [see  (\ref{dscr}) below] \cite{Dsc} of 
the characteristic determinant $\Delta$ 
is {\em negative} (see section \ref{sec:ts}). 
Indeed, general solutions in the parameter range $D> 0$ 
are {\em nonoscillating} and can be
represented as a sum of three exponentially damped terms, 
wherein each term corresponds to a real
cubic root of the characteristic determinant $\Delta$ 
(see section \ref{sec:ts}).

Surprisingly enough, the limitation of Torrey's 
general solution (\ref{trsl}) 
to the parameter range $D<0$ and the unawareness of 
{\em Cardano's formula} had been since without exception 
repeated in the literature \cite{Tr,WLK} 
and in textbooks (cf section 3.5 of Ref. \cite{AE}) for several decades.
The latter is quite remarkable given the role that 
Torrey's analytic solutions played 
in quantum optics and magnetic resonance literature \cite{WLK,AE}.
It took several decades before Hore and McLauchlan 
had finally made use of Cardano's formula \cite{Cf} 
and formally determined the cubic roots \cite{HM}.
However, their analytic solution has other deficiencies to
be discussed in section \ref{sec:disc}.

In what follows, we closely follow the original Torrey's 
approach \cite{Tr,HM,MK} and supplement it with two additional
elementary tools:
\begin{itemize}

\item (i) the partial fraction expansion, or 
{\em partial fraction decomposition} (PFD) \cite{Pfdw} over
the field of {\em real} numbers
(see Appendix \ref{sc:pfe}), combined with the classification
of roots of a real cubic polynomial, and 

\item (ii) a fully compensated formula for real roots of a real
cubic polynomial [see (\ref{crdzl}) below], 
which has recently been derived \cite{AMr} from 
{\em Cardano's formula} \cite{Cf}. 

\end{itemize}
The first tool will enable one to classify and, up to an explicit
knowledge of the roots of $\Delta$, 
determine all the possible functional forms of 
solutions of the Bloch equations (\ref{bleq})
for all values of $D\lesseqqgtr 0$.
The second tool will allow one to 
determine the roots of $\Delta$, and hence
solutions of the Bloch equations, explicitly.

The outline of the article is as follows.
In section \ref{sec:nota} we
 summarize notation and give some basic definitions.
In section \ref{sec:mt} a full advantage of the
partial fraction decomposition over real number field is made, 
all possible solution types (e.g. of non Torrey type) are classified 
[see (\ref{lplsf})]. In section \ref{sec:uso} 
the parameter range of unorthodox solutions is determined.
In section \ref{sec:ts} decay constants corresponding  
to each solution type are determined by Cardano's formula.
In section \ref{sec:dgr} sufficient and necessary conditions for
{\em doubly} and {\em triply} degenerate real
roots are established in terms of the model parameters.
Steady state solutions and remaining numerical 
constants of different solution 
types [see (\ref{lplsf})] are determined in section \ref{sec:prp}.
In section \ref{sec:disc} connection to earlier results 
and various conditions of the
applicability of our results are discussed. 
We then conclude by section \ref{sec:conc}. A number of 
formulae and intermediary steps are relegated 
to Appendices \ref{sc:pfe} and \ref{ap:iltf}.

\section{Bloch equations}
\label{sec:nota}
The Bloch equations are a linear system of 
differential equations with {\em constant coefficients} 
for a {\em Bloch vector} with components $(u,v,w)$,
also known as the {\em Bloch variables} 
\cite{Bl,Tr,HM,MK,McH,WLK,AE,Ya},
\begin{eqnarray}
u' &=& -\gamma_t u - \delta v,
\nonumber\\
v' &=& -\gamma_t v + \delta u  + \Omega\, w,
\nonumber\\
w' &=&  -\gamma w - \Omega v +\gamma w_{eq},
\label{bleq}
\end{eqnarray}
where the prime denotes derivative with respect to a 
parameter $t$ defined in  Table 1, 
and $w_{eq}$ is an equilibrium value of $w$. 
Occasionally we denote the Bloch variables as $(u,v,w)=(u_1,u_2,u_3)$.
\vspace*{0.3cm}

{\bf Table 1.} The Bloch equations parameters in the case of a two-level
system and in the case of magnetic resonance (MR). 

\begin{center} 
\begin{tabular} {c|cc} \hline  &  two-level system &  MR  
\\ 
\hline 
 $t$  &  time $\tau$ &  rescaled time  $g H_1 \tau$
\\ 
 $\gamma$  &  $1/T_1$   & $1/(g H_1 T_1)$ 
\\
 $\gamma_t$  &  $1/T_2$   & $1/(g H_1 T_2)$
\\
 $\omega_0$  &  $(\epsilon_2-\epsilon_1)/\hbar$   & Larmor frequency  $g H_0$  
 \\ 
 $\delta$  &  $\omega_0-\omega$   & $(\omega_0-\omega)/(g H_1)$ 
  \\ 
 $\Omega$  &  $\frac{|{\mbox{\boldmath $\mu$}}\cdot{\bf E}_0|}{\hbar}$   &  $-1$
 \\
 \hline
\end{tabular}
\end{center}
\vspace*{0.3cm} 
%
The meaning of the parameters depends on the problem solved.
In the case of a two-level
system with energies $\epsilon_1$ and 
$\epsilon_2>\epsilon_1$ interacting with 
time-harmonic perturbing classical field $E(t)=E_0 \cos\omega t$
with the driving frequency $\omega$ and amplitude $E_0$, 
the {\em Bloch variables} $u,\, v,\, w$ 
are related to the elements of a $2\times 2$ hermitian 
density matrix $\rho_{ij}=\rho_{ji}^*$ \cite{WLK,Ya}
\begin{equation}
u =2\mbox{Re } \rho_{12},~~~
  v= 2\mbox{Im } \rho_{12},~~~
             w=  \rho_{22}-\rho_{11}.
\label{uvwdf}
\end{equation}
The parameter $\delta$ stands for the driving field 
{\em detuning} from the intrinsic
resonance frequency $\omega_0$, whereas
$\Omega$, which reduces to the Rabi frequency at 
zero detuning,
accounts for the interaction strength - it is 
entirely determined by the coupling with the driving field
(e.g. between the 
electric field amplitude and the transition dipole
moment ${\mbox{\boldmath $\mu$}}$) (see  Table 1).
Phenomenological constants 
$\gamma$ and $\gamma_t$ describe 
decay rates of the excited state (the total
population relaxation rate) and of the coherence, respectively, where
$T_1$ is the {\em spontaneous emission} lifetime and 
$T_2$ (typically $\geq T_1$) is the total {\em dephasing} time \cite{AE,Ya}.
According to (\ref{uvwdf}), $w$ is the single atom 
population difference, also called
{\em inversion} \cite{AE,Ya}. The condition $w(0)= -1$, which
(together with $u_0=v_0=0$) is often taken 
as the initial condition \cite{AE,MK,NJ},
means that the two-level atom is initially in its ground state. 
The third equation in (\ref{bleq}) shows that
$v$ is the component effective in coupling to the field to produce
energy changes. Thus $v$ determines the {\em absorptive} 
({\em in-quadrature} with the field ${\bf E}$)
component and $u$ the {\em dispersive} ({\em in-phase}) 
component of the atomic transition dipole moment. 
Note in passing that $w_{eq}$ is not necessary 
a thermal equilibrium value, since some pump mechanism may 
be present that causes $w_{eq}$ at equilibrium to have 
some fixed value that is 
different from its thermal equilibrium value \cite{Ya}.

In the theory of magnetic resonance of precessing (nuclear or electronic)
spins \cite{Bl,Tr,HM,MK}, the time constants $T_2$ and $T_1$ 
correspond to the {\em longitudinal} 
and {\em transverse} relaxation constants \cite{Bl,Tr,HM,MK} and, in contrast
to the optics,
$T_2$ is always less than or equal to $T_1$ \cite{Hr}.
Assuming a constant magnetic field ${\bf H}_0$ applied
along the $z$-axis and a time-harmonic component 
$2{\bf H}_1 \cos\omega t$
applied in the $x$-direction, the components $u,v,w$ in Eqs.
(\ref{bleq}) are formed by the respective
$x$, $y$, $z$ components of a nuclear or electron (macroscopic)
magnetization ${\bf M}$, and $w_{eq}$ is related to the 
static magnetization ${\bf M}_0=\chi_0 {\bf H}_0$, where $\chi_0$
is the static susceptibility \cite{Bl,Tr}. 
Eqs. (\ref{bleq}) are then valid with
the time $\tau$ together with the homogeneous lifetimes $T_1$ and $T_2$ 
being rescaled by $g H_1$,
where $g$ is the absolute value of the gyromagnetic ratio (see  Table 1).

Importantly, in any case is the ratio  $\Omega/|\gamma-\gamma_t|$ proportional to the 
driving field (either ${\bf E}_0$ or ${\bf H}_1$) and (see Table 1)
\begin{equation}
\frac{\delta}{|\gamma-\gamma_t|} = \frac{\omega_0-\omega}{\left| \frac{1}{T_1}- \frac{1}{T_2} \right|} \cdot
\label{odd}
\end{equation}

\section{Classification of possible solution types}
\label{sec:mt}
Following Torrey's approach \cite{Tr,HM,MK}, 
upon applying the Laplace transform, 
\begin{equation}
\tilde{f} (p) = \int_0^\infty f(t) e^{-pt} dt,
\end{equation}
the Bloch equations (\ref{bleq}) are transformed 
into the matrix equation
\begin{equation}
\left(
\begin{array}{ccc}
p+\gamma_t & \delta & 0 
\\
-\delta &  p+\gamma_t & -\Omega
\\
0 & \Omega &  p+\gamma
\\
\end{array}\right)
\left(
\begin{array}{c}
\tilde{u} 
\\
\tilde{v} 
\\
\tilde{w} 
\end{array}\right)
=
\left(
\begin{array}{c}
u_0 
\\
v_0
\\
W_0
\end{array}\right),
\label{chm}
\end{equation}
where tilde denotes the Laplace transform
of the Bloch variables,
\begin{equation}
W_0 = w_0 +\frac{\gamma w_{eq}}{p},
\end{equation}
and $u_0$, $v_0$, $w_0$ are the initial values of 
the Bloch variables.
Cramer's rule then yields
\begin{equation}
\tilde{u}_l = \frac{f_l(p)}{\Delta(p)},
\label{lpls}
\end{equation}
where $\tilde{u}_l$, $l=1,2,3$, stands for $\tilde{u}$, $\tilde{v}$, and $\tilde{w}$, 
respectively,  and
\begin{equation}
f_l (p) =\left\{
\begin{array}{cc} 
 u_0[(p+\gamma_t)(p+\gamma)+\Omega^2] - v_0\delta(p+\gamma) - W_0 \delta \Omega, & l=1
\\
u_0\delta(p+\gamma) + v_0(p+\gamma_t)(p+\gamma) + W_0 \Omega\, (p+\gamma_t), & l=2
\\
 -u_0 \delta \Omega -v_0\Omega(p+\gamma_t) + W_0 [(p+\gamma_t)^2+\delta^2],  & l=3
\end{array}\right.
\label{fld}
\end{equation}
The determinant $\Delta(p)$ of the coefficient matrix 
is a {\em real} cubic polynomial 
\begin{equation}
\Delta(p)=p^3 + a_2 p^2 + a_1 p + a_0,
\label{dtp}
\end{equation}
where
\begin{eqnarray}
a_2 &=& \gamma+2\gamma_t,
\label{a2}
\\
a_1 &=& \gamma_t^2 +2\gamma\gamma_t+\delta^2+\Omega^2,
\label{a1}
\\
a_0 &=& \gamma\gamma_t^2+\gamma\delta^2+\gamma_t\Omega^2.
\label{a0}
\end{eqnarray}

The multiplication of both the numerator and denominator
on the right-hand side of (\ref{lpls}) by $p$ changes each
 $f_l(p)$ defined by Eqs. (\ref{fld})
into a cubic polynomial, whereas the 
ratio on the right-hand side of (\ref{lpls}) becomes the ratio 
of a cubic and quartic polynomials. 
Given that $\Delta(p)$ is a {\em real} cubic polynomial
[see (\ref{a2}) to (\ref{a0}) 
for the coefficients $a_j$ in (\ref{dtp})],
this suggests the application of the PFD over 
the field of {\em real} numbers
[see  (\ref{pfd}) in Appendix \ref{sc:pfe}]. 
The PFD enables one to express the fractions such as that in (\ref{lpls}) 
as a sum of much simpler fractions, whose inverse Laplace transform 
may be readily available.
To this end note [e.g. by the fundamental theorem of algebra (\ref{fta})] 
that as any {\em real} cubic polynomial, $\Delta(p)$ 
can have either (i) one real root and a pair of 
complex conjugate roots or (ii) three real roots.
Given that all the coefficients $a_j$ in (\ref{dtp}) 
are {\em positive real} numbers, one can prove 
additionally that:
\begin{itemize}

\item $\Delta(p)$ has always at least one 
{\em negative real} root ({\bf P1});
 
\item if $\Delta(p)$ has three {\em real} roots, 
they are all {\em negative} ({\bf P2});

\item if not all roots of $\Delta(p)$ are {\em real} than there is 
{\em one} {\em real} root and {\em two} 
complex conjugate (c.c.) roots ({\bf P3});

\item if at least {\em two} roots of $\Delta(p)$ coincide, 
they are all {\em negative real} roots ({\bf P4}).
It may be that $\Delta(p)$ has a {\em double} real root and 
another distinct single real root; alternatively, all 
three roots of $\Delta(p)$ coincide yielding a {\em triple} real root.

\end{itemize}
The properties {\bf P1} and {\bf P2} can be shown 
to be a straightforward consequence of the 
{\em intermediate value theorem} when applied to a cubic
polynomial $\Delta(p)$ with {\em positive real}
coefficients $a_j$.
Obviously, a sufficient condition for {\bf P1} is $\Delta(0) = a_0 > 0$, 
whereas a sufficient condition for {\bf P2} is 
\begin{equation}
\Delta(p) > 0 \hspace*{4cm} (p\ge 0).
\label{dtpp}
\end{equation}
The property {\bf P3} follows from {\bf P1} and 
the fundamental theorem of algebra (\ref{fta}) 
applied to a {\em real} cubic polynomial.
Eventually, the property {\bf P4} follows 
upon combining {\bf P1} to {\bf P3}. 
Note in passing that 
we disregarded a special case of $a_0=0$, 
which is treated at the end of section \ref{sec:prp}.
To this case belongs also the trivial undamped 
case $\gamma=\gamma_t=0$ (which implies both $a_0=0$ and $a_2=0$),
in which case the Bloch equations (\ref{bleq}) describe the
precession of a classical gyromagnetic moment 
in a magnetic field \cite{FVH,AE}.

Given the properties {\bf P1} to {\bf P4},
and on writing {\em negative} real roots of $\Delta(p)$ as
$x_j=-\kappa_j$, $j=1,2,3$ ($\kappa_j>0$),
the application of the PFD [see  (\ref{pfd}) in Appendix \ref{sc:pfe}]
enables one to decompose each $f_l(p)/\Delta(p)$ as
\begin{equation}
\frac{f(p)}{\Delta(p)} = \frac{pf(p)}{p\Delta(p)} =
\left\{
\begin{array}{ll}
 \frac{A_0}{p} + \frac{A_1}{p+\kappa_1}+\frac{A_2(p+b) +A_3}{(p+b)^2 + s^2},
& \mbox{(pair of c.c. roots)}
\\
 \frac{A_0}{p} + \frac{A_1}{p+\kappa_1}+\frac{A_2}{p+\kappa_2}+\frac{A_3}{p+\kappa_3},
& \mbox{(distinct real roots)} 
\\
 \frac{A_0}{p} + \frac{A_1}{p+\kappa_1}+\frac{A_2}{p+\kappa_2}+\frac{A_3}{(p+\kappa_2)^2},
& \mbox{(double real root)}
\\
 \frac{A_0}{p} + \frac{A_1}{p+\kappa_1}+\frac{A_2}{(p+\kappa_1)^2}+\frac{A_3}{(p+\kappa_1)^3},
&\mbox{ (triple real  root)}.
\end{array}\right.
\label{tdc}
\end{equation}
Herein and below the index $l$ labeling different 
Bloch variables will be suppressed 
unless explicitly required.
In the first line of Eqs. (\ref{tdc}), corresponding to cubic roots
$-\kappa_1$, $z$, and $\bar{z}$, we have recast 
the irreducible quadratic factor $(p-z)(p-\bar{z})$ 
in Torrey's form $(p+b)^2 + s^2$.
Given that the coefficients $a_2$ and $a_0$ 
in Eqs. (\ref{a2}) and (\ref{a0}) can
be alternatively expressed in terms of cubic roots 
$z_1$, $z_2$, and $z_3$ of $\Delta(p)$
as $a_2=-(z_1+z_2+z_3)$ and $a_0=-z_1z_2z_3$,
the Torrey constants $b$ and $s$ 
can be entirely expressed in terms of $\kappa_1$, $a_2$ and $a_0$,
\begin{equation}
b=-\mbox{Re }z=\frac{a_2-\kappa_1}{2},~~~~~
s=\sqrt{|z|^2-b^2}=\sqrt{\frac{a_0}{\kappa_1}-b^2}.
\label{bsd}
\end{equation}
On substituting (\ref{tdc}) back into (\ref{lpls}) one can perform
the inverse Laplace transform (see Appendix \ref{ap:iltf}). 
Note in passing that the property {\bf P2} additionally
guarantees the very existence of the inverse Laplace transform. 
(Indeed, if one of the real roots were positive, one would
face singular integrals.) After the inverse Laplace transform
one arrives at the following {\em complete} set of 
the possible functional forms of solutions of the 
Bloch equations (\ref{bleq}),
\begin{equation}
u_l(t) = 
\left\{
\begin{array}{ll}
A_0 + A_1 e^{-\kappa_1t} + A_2e^{-bt}\cos(st) + (A_3/s) e^{-bt}\sin(st), & 
\mbox{(pair of c.c. roots)}
\\
A_0 + A_1 e^{-\kappa_1 t} + A_2e^{-\kappa_2 t} + A_3  e^{-\kappa_3 t}, & \mbox{(distinct real roots)} 
\\
A_0 + A_1 e^{-\kappa_1 t} + A_2e^{-\kappa_2 t} + A_3 t e^{-\kappa_2 t},  & \mbox{ (double root)}
\\
A_0 + A_1 e^{-\kappa_1 t} + A_2 t e^{-\kappa_1 t} 
            + \frac{1}{2} A_3 t^2 e^{-\kappa_1 t}, & \mbox{ (triple root)}.
\end{array} \right.
\label{lplsf}
\end{equation}
The PFD's given by Eqs. (\ref{tdc}) exhaust all 
the possible PFD's of the ratio $f(p)/\Delta(p)$.
There is {\em no other decomposition possible}.
Only the first line in Eqs. (\ref{tdc}) and (\ref{lplsf}) 
corresponds to the Torrey solution \cite{Tr},
whereas remaining lines yield {\em unorthodox} (i.e. non Torrey) solutions.
Importantly, the PFD guarantees that the respective sets of numerical 
constants $A_0$, $A_1$, $A_2$, and $A_3$
for different Bloch variables, together with
constants $b$ and $s$, are all {\em real} numbers.

\section{Parameter range of unorthodox solutions}
\label{sec:uso}
To this end, we have not verified yet if any of the above solution
types could be attained within the physical range of parameters in the 
Bloch equations (\ref{bleq}). In order to proceed,
one needs to determine
the value of the discriminant $D$ of $\Delta(p)$. According to \cite{AMr}
\begin{eqnarray}
D &=& a_1^2a_2^2 - 4a_1^3 - 4a_0a_2^3 -27a_0^2+ 18 a_0a_1a_2
\nonumber\\
&=& (\gamma-\gamma_t)^2 \Omega^2 \left( \Omega^2 + 20 \delta^2 \right) 
- 4 (\gamma-\gamma_t)^2 \delta^2 \left[(\gamma-\gamma_t)^2 + 2\delta^2 \right]
\nonumber\\
&&
        - 4(\delta^2 + \Omega^2)^3.
\label{dscr}
\end{eqnarray}
In terms of the roots, the discriminant of a general polynomial of 
the $n$th order is given by \cite{Dsc}
\begin{equation}
    D=a_n^{2n-2}\prod_{i<j}{(z_i-z_j)^2},
\label{vdmd}
\end{equation}
where $a_n$ is the leading coefficient [$a_n=a_3=1$ in (\ref{dtp})] 
and $z_1,\ldots,z_n$ 
are the roots (counting multiplicity) of the polynomial. 
For a cubic polynomial with {\em real} coefficients, the use of 
the general form of the discriminant \cite{Dsc} together 
with its {\em real} value [see (\ref{dscr})] 
enables one to relate the nature of roots to the value
of $D$ as follows:
\begin{itemize}

\item $D < 0$: a cubic polynomial has {\em one} {\em real} 
root and {\em two} complex conjugate roots;

\item $D > 0$: there are {\em three} distinct {\em real} roots;

\item $D = 0$: at least {\em two} roots coincide, 
and they are all {\em real}.

\end{itemize}
To this end one can show that Torrey special solutions labeled
by  {\bf T1} and {\bf T3} introduced in section \ref{sc:intr}
are always described by {\em exponentially damped harmonic oscillations}.
Indeed, in the particular case of {\em strong collisions} ({\bf T1}),
Eq. (\ref{dscr}) reduces for $\gamma=\gamma_t$ to
\begin{equation}
D = - 4(\delta^2 + \Omega^2)^3 < 0.
\label{dscrsc}
\end{equation}
In the particular case of {\em intense external field} ({\bf T3}),
Eq. (\ref{dscr}) reduces for $\Omega \gg \gamma_t,\,\gamma$ to
\begin{equation}
D = - \Omega^4 \left[4 \Omega^2 - (\gamma-\gamma_t)^2 \right]< 0.
\label{dscrzsf}
\end{equation}

Provided that $\gamma\ne \gamma_t$, Table 1 and ensuing discussion 
at the end of section \ref{sec:nota} suggest to
measure $\delta^2$ and $\Omega^2$ in the units of $(\gamma-\gamma_t)^2$. 
Upon introducing $\alpha\geq 0$ and $\beta\geq 0$ through
\begin{equation}
\delta^2=\alpha (\gamma-\gamma_t)^2, \hspace*{1.8cm}
\Omega^2=\beta (\gamma-\gamma_t)^2,
\label{alb}
\end{equation}
Eq. (\ref{dscr}) becomes
\begin{eqnarray}
D &=& - 4(\gamma-\gamma_t)^6\, \left[
\beta^3 +\beta^2\left(3\alpha - \frac{1}{4}\right) 
            + \beta\alpha(3\alpha-5) +\alpha(\alpha+1)^2 \right]
\nonumber\\
&=& - 4(\gamma-\gamma_t)^6\, h(\alpha,\beta).
\label{dscra}
\end{eqnarray}
In the particular case of a {\em zero detuning}, $\delta=\alpha=0$ 
(Torrey's case  {\bf T2}), 
the discriminant is greater than or equal to zero and all roots are real
provided that $\beta\leq 1/4$, or
\begin{equation}
\Omega \leq \frac{|\gamma_t-\gamma|}{2}\cdot
\label{erc}
\end{equation}
Obviously, each of the regions $D\lesseqqgtr 0$ can be attained by a 
suitable choice of physical parameters 
(see figures \ref{fgdabz} and \ref{fgdab}) in Torrey's case {\bf T2}.
Note in passing that $h(\alpha,\beta)$
considered as the cubic polynomial in $\beta$ has 
 only {\em positive} coefficients for $\alpha\geq 5/3$. Thus 
any  real root of $h$ has to be {\em negative}. Consequently 
$D(\alpha,\beta)<0$ for any $\beta\geq 0$ and $\alpha\geq 5/3$.
Actually a much stronger statement can be proven: 
\begin{equation}
D(\alpha,\beta)<0 \hspace*{1.2cm} \mbox{for any} \hspace*{0.4cm} \beta\geq 0
\hspace*{0.4cm} \mbox{and}\hspace*{0.4cm} \alpha>1/27.
\end{equation}
The proof proceeds as follows.
Considered as a cubic polynomial in $\beta$, one has 
$h(\alpha,0)>0$ for any $\alpha>0$. Thus, as a straightforward consequence of the 
{\em intermediate value theorem}, $h(\alpha,\beta)$ possesses a {\em negative 
real root} for any $\alpha>0$.
Now it suffices to show that the discriminant $d_h$ of $h(\alpha,\beta)$ 
considered as a cubic polynomial in $\beta$ is {\em negative}. 
This is indeed the case.
Upon tedious but straightforward calculations one finds
\begin{equation}
d_h=-\frac{1}{16}\, \alpha (27\alpha -1)^3
\left\{
\begin{array}{cl}
<0, & \alpha>1/27,
\\
>0, & 0<\alpha<1/27.
\end{array}\right.
\label{pdc}
\end{equation}
The first inequality prohibits any additional real root of $h(\alpha,\beta)$,
and hence any root for $\beta\in (0,\infty)$ and $\alpha>1/27$. 
Because $h(\alpha,0)>0$ and $h(\alpha,\beta\rightarrow\infty)\rightarrow +\infty$, 
one has necessarily $h(\alpha,\beta)>0$ for $\beta\in(0,\infty)$.
The absence of zeros of $D(\alpha,\beta)$ for $\alpha>1/27$
is demonstrated in figures \ref{fgdabz} and \ref{fgdab}, which plot 
the functional dependence of 
$D_c(\alpha,\beta) = -D/108$ [see \ref{ddf})] on $\beta$ for selected values
of $\alpha$. 
Figure \ref{fgbnd} shows the boundary of the $D_c< 0$ region 
in the $(\Omega^2,\delta^2)$ plane.
The boundary is exactly described by \cite{NJ}
\begin{equation}
\alpha=-\frac{1}{3}\, (2+3\beta) +  \frac{2}{3}\, \sqrt{1+27\beta}\,\cos(\theta+\theta_0),
\end{equation}
where
\begin{equation}
\theta= \frac{1}{3}\, \cos^{-1}\left[\frac{8-27\beta(20+27\beta)}{8(1+27\beta)^{3/2}}
\right],
\end{equation}
and $\theta_0=0$ for the part of the boundary between the origin and 
the cusp point, whereas $\theta_0=4\pi/3$ for the part of the boundary between 
the cusp point and $\beta=1/4$.
As it will be demonstrated in section
\ref{sec:dgr}, the critical value $\alpha= 1/27$, which corresponds to the cusp
point in figure \ref{fgbnd},
is related to the sufficient and necessary 
conditions (\ref{trc}) for the occurence of a {\em triply} 
degenerate real root.
Since the conditions for a {\em triply} 
degenerate real root 
can also be satisfied (see section \ref{sec:dgr} below), all 
functional forms of solutions of the 
Bloch equations listed in Eqs. (\ref{lplsf}) are physically 
achievable.

\section{Decay constants}
\label{sec:ts}
In this section, various parameter ranges corresponding
to each of the solutions in (\ref{lplsf}) 
of the Bloch equations (\ref{bleq}) are identified 
and decay constants of various solution types 
are explicitly determined. 
A prerequisite for that is a relation between roots of 
the cubic polynomial $\Delta(p)$ [see (\ref{dtp})] and the
physical parameters of the 
Bloch equations (\ref{bleq}). The latter is provided
by means of  {\em Cardano's} formula \cite{Cf,Cr}.
{\em Cardano's} roots $z_l$ of a 
cubic polynomial (\ref{dtp}) have conventionally been written
as follows \cite{Cf}:
\begin{eqnarray}
z_1 &=& -\frac{1}{3}\, a_2 + (S_+ + S_-),	
\nonumber\\
z_2 &=& -\frac{1}{3}\, a_2-\frac{1}{2}\, (S_+ + S_-)
             +\frac{1}{2}\, i\sqrt{3}\,(S_+ - S_-),	
\nonumber\\
z_3 &=& -\frac{1}{3}\, a_2 -\frac{1}{2}\, (S_+ + S_-)
              -\frac{1}{2}\, i\sqrt{3}\,(S_+ - S_-),
\label{crdf}	
\end{eqnarray} 
where
\begin{equation}	
S_+ = \sqrt[3]{R + \sqrt{D_c}},
\hspace*{2cm}
S_- = \sqrt[3]{R - \sqrt{D_c}},
\label{sdf}
\end{equation}
and, on substituting the values of the 
coefficients $a_j$ according to  Eqs. (\ref{a2})-(\ref{a0}),
\begin{eqnarray}
R &=& \frac{9a_2a_1-27a_0-2a_2^3}{54} 
  = (\gamma-\gamma_t)\left[ \frac{1}{6}\, (\Omega^2 - 2\delta^2) -\frac{1}{27}\, (\gamma-\gamma_t)^2    
\right],
\label{rdf}
\\
Q &=& \frac{3a_1-a_2^2}{9} = \frac{1}{3}\,(\delta^2+\Omega^2)
 -\frac{1}{9}\, (\gamma-\gamma_t)^2,
\label{qdf}
\\
D_c &=& Q^3+R^2 = -\frac{D}{108} =  \frac{1}{108} \times
\nonumber\\
&&
\left[
4 (\gamma-\gamma_t)^4\delta^2 + 4(\delta^2 + \Omega^2)^3 
  - (\gamma-\gamma_t)^2\left( \Omega^4 + 20  \delta^2\Omega^2 
- 8  \delta^4\right)\right].\phantom{xxxx}
\label{ddf}
\end{eqnarray}
Note in passing that $\kappa_1$ in (\ref{lplsf}) can always be obtained 
from the first Cardano's root (\ref{crdf}), 
\begin{equation}
\kappa_1 = \frac{1}{3}\, a_2 - \sqrt[3]{R + \frac{\sqrt{-D}}{6\sqrt{3}}}
-\sqrt[3]{R - \frac{\sqrt{-D}}{6\sqrt{3}}},
\label{al1d}	
\end{equation}
provided that (i) a cubic root $\sqrt[3]{x}$ of a real 
number $x$ is chosen to be a real number with the same sign
as $x$ ({\bf R1}); 
(ii) the respective cubic roots $\sqrt[3]{x\pm iy}$ of 
complex conjugate numbers $x\pm iy$ remain
 complex conjugate numbers ({\bf R2}) \cite{AMr}.
As an example, in the particular case of {\em strong collisions} ({\bf T1}),
one has $a_2=3\gamma$, $R=0$ [see (\ref{rdf})],
\begin{equation}
S_\pm = \sqrt[3]{\pm \sqrt{D_c}} =\pm \sqrt{\sqrt[3]{D_c}}
= \pm \frac{1}{\sqrt{3}}\, \sqrt{\delta^2 + \Omega^2}
\end{equation}
[see (\ref{sdf}) combined with that $D_c=-D/108$], 
and the Cardano's formula (\ref{crdf}) 
reproduces the known roots \cite{Tr,AE}
\begin{equation}
x_1 = -\gamma, ~~~~~ z_{2,3}=-\gamma \pm i\sqrt{\delta^2 + \Omega^2}.
\end{equation}
For $\gamma\ne \gamma_t$ one finds
\begin{equation}	
S_\pm = (\gamma-\gamma_t) \sqrt[3]{\frac{1}{6}\, (\beta-2\alpha) - \frac{1}{27} 
\pm \frac{1}{3\sqrt{3}} \sqrt{h(\alpha,\beta)} }.
\label{sdfa}
\end{equation} 
The real roots could be explicitly determined 
according to formula \cite{AMr}
\begin{equation}
x_{l+1}= -\frac{1}{3}\, a_2 + 2 \mbox{Re }\left(e^{i 2\pi l/3} S \right).
\label{crdzl}	
\end{equation} 
The root formula (\ref{crdzl}) is fully {\em compensated}: it does 
not matter either which of $S_+$ and $S_-$ has been taken for $S$, 
or which of the cubic roots has initially been taken for 
$S_\pm$ in (\ref{sdf}). The set of real cubic roots given 
by (\ref{crdzl}) remains invariant under any of the above choices, 
with a particular choice affecting only an irrelevant root 
permutation within the root set. 

In the particular case of a {\em zero detuning}, $\delta=0$, 
or {\em exact resonance} (Torrey's case {\bf T2}), 
$\Delta(p)$ factorizes into the product [see (\ref{chm})]
\begin{equation}
\Delta(p)=(p+\gamma_t)[(p+\gamma_t)(p+\gamma)+\Omega^2].
\end{equation}
Thereby one needs only to solve a quadratic equation to obtain
the roots (Appendix of Ref. \cite{Tr}; section 3.5 of Ref. \cite{AE})
\begin{equation}
x_1=- \gamma_t,~~~~~
         z_{2,3} =-\frac{\gamma+\gamma_t}{2} \pm \sqrt{(\gamma-\gamma_t)^2-4\Omega^2}.
\label{aeroot}
\end{equation}

\section{Degenerate real roots}
\label{sec:dgr}
In virtue of the  definition (\ref{ddf}), 
a necessary and sufficient
condition for $D>0$ (or equivalently $D_c < 0$), i.e., for
purely damped {\em nonoscillating} solutions, is 
\begin{equation}
Q^3 < -R^2  < 0,
\end{equation}
which in turn requires $Q< 0$. Given (\ref{qdf}), the latter implies 
\begin{equation}
\delta^2+\Omega^2 < \frac{1}{3}\, (\gamma_t -\gamma)^2
\Longleftrightarrow \alpha+\beta < \frac{1}{3} ,
\label{qlzc}
\end{equation}
which is necessary (but not sufficient) condition for $D>0$.
Obviously, the condition (\ref{qlzc}) cannot be satisfied 
in a nonzero driving field for $\gamma=\gamma_t$.

One has $D=0$ [or $D_c=0$ - see (\ref{ddf})] if 
\begin{equation}
Q^3 = -R^2 \leq 0.
\label{q3r2}
\end{equation} 
Then according to (\ref{sdf}) 
\begin{equation}
S_\pm =\sqrt[3]{R},\hspace*{3cm}(D_c=0).
\label{ddr}
\end{equation}
Provided that the second inequality in (\ref{q3r2}) is a sharp 
{\em inequality}, one has a {\em doubly} degenerate real root. 
Indeed, according to formula (\ref{crdzl}),
$\Delta(p)$ has exactly two degenerate roots $x_{2,3}=z_{2,3}$ 
if and only if $D=0$ and $R\ne 0$, and
consequently, in virtue of the definition (\ref{ddf}), $Q< 0$.
A necessary condition for two degenerate roots 
is again given by  (\ref{qlzc}).
As an illustration, in the special case of $\delta=0$ ({\bf T2}) 
the case of $D=0$ corresponds to $\Omega=|\gamma_t-\gamma|/2$ 
[see (\ref{erc})] and $R = (\gamma-\gamma_t)^3/216$,
which, according to our root convention {\bf R1} and {\bf R2}, 
yields $S_\pm = \sqrt[3]{R} = (\gamma -\gamma_t)/6$ [see (\ref{ddr})]
\begin{equation}
x_{2,3} =-\frac{\gamma+\gamma_t}{2},
\label{ddgr}
\end{equation} 
and hence a {\em doubly} degenerate real root [see (\ref{aeroot})].

The conditions for a {\em triply} degenerate real root
can be satisfied for a nonzero $\Omega$ only with a nonzero detuning. 
Indeed, according to (\ref{crdzl}), 
a cubic polynomial can have a triply degenerate real root
if and only if $S_\pm =0$, in which case the rotating term 
in (\ref{crdzl}) vanishes, and 
\begin{equation}
x_{1,2,3} = -\frac{1}{3}\, a_2= -\frac{1}{3}\,(\gamma+2\gamma_t)<0.	
\label{tdr}
\end{equation}
According to  (\ref{sdf}), $S_\pm =0$ is possible 
if, in addition to $D_c=0$, one has also $R=0$. 
The above conditions can only be satisfied if $R=Q=0$ 
[cf the first equality in (\ref{ddf})], or,
given Eqs. (\ref{rdf}) and (\ref{qdf}), when simultaneously
\begin{eqnarray}
2 (\gamma-\gamma_t)^2 &=& 9 \Omega^2 - 18 \delta^2,
\label{rdfz}
\\
(\gamma-\gamma_t)^2 &=& 3 (\delta^2+\Omega^2).
\label{qdfz}	
\end{eqnarray}
In term of $\delta$ and $\Omega$
\begin{eqnarray}
\delta^2 &=& \frac{1}{27}\, (\gamma-\gamma_t)^2 \Longleftrightarrow \alpha= \frac{1}{27},
\\
\Omega^2 &=& 8  \delta^2 = \frac{8}{27}\, (\gamma-\gamma_t)^2 
\Longleftrightarrow \beta= \frac{8}{27} ,
\label{trc}
\end{eqnarray} 
and one finds $\alpha+\beta=1/3$ [see (\ref{qlzc})].
Thus unless $\gamma=\gamma_t$ ({\bf T1}) one can always attain
the case of a triply degenerate real root for $\Omega\neq 0$.

\begin{figure}[tbp]
\begin{center}
\epsfig{file=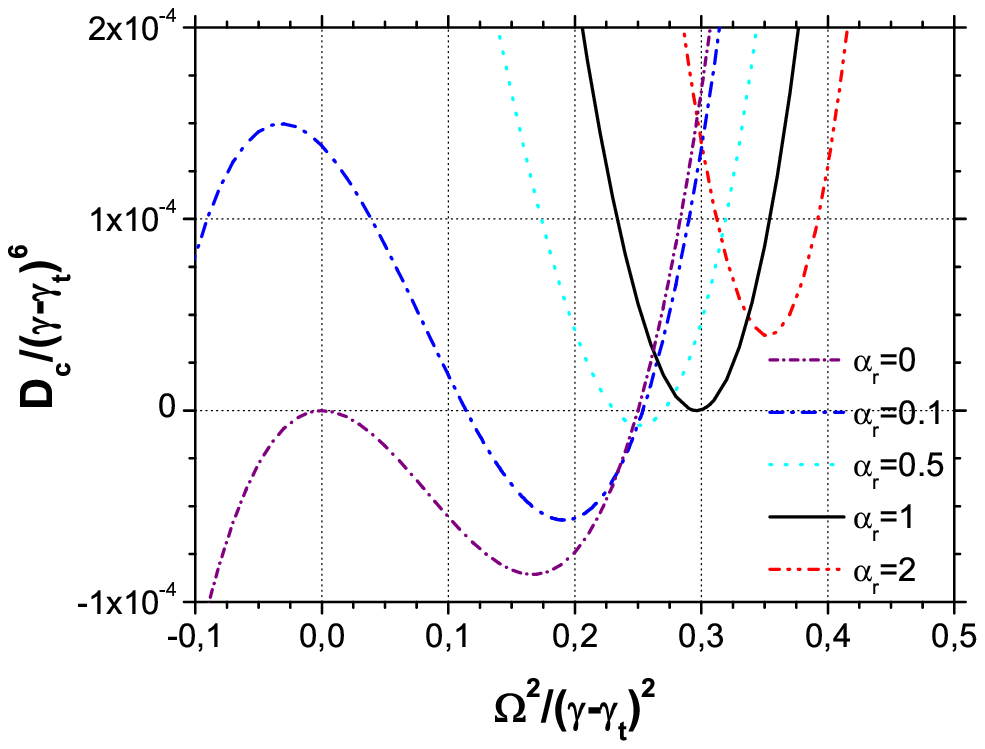,width=18cm,clip=0,angle=0}
\end{center}
\caption{}
\label{fgdabz}
\end{figure}

\begin{figure}[tbp]
\begin{center}
\epsfig{file=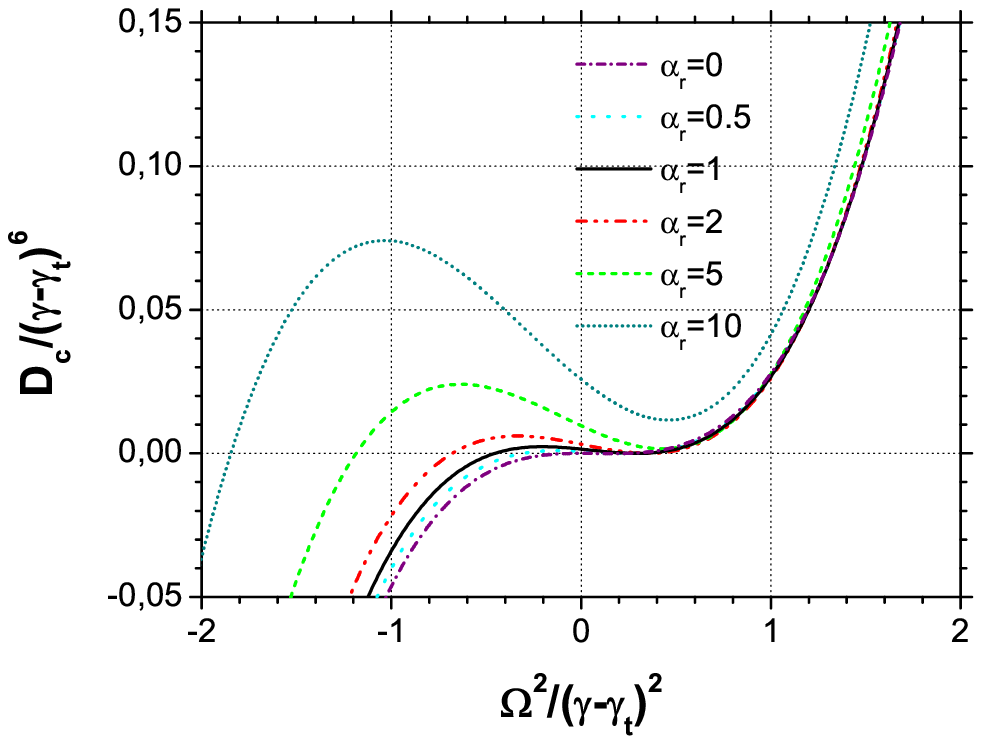,width=18cm,clip=0,angle=0}
\end{center}
\caption{}
\label{fgdab}
\end{figure}

\begin{figure}[tbp]
\begin{center}
\epsfig{file=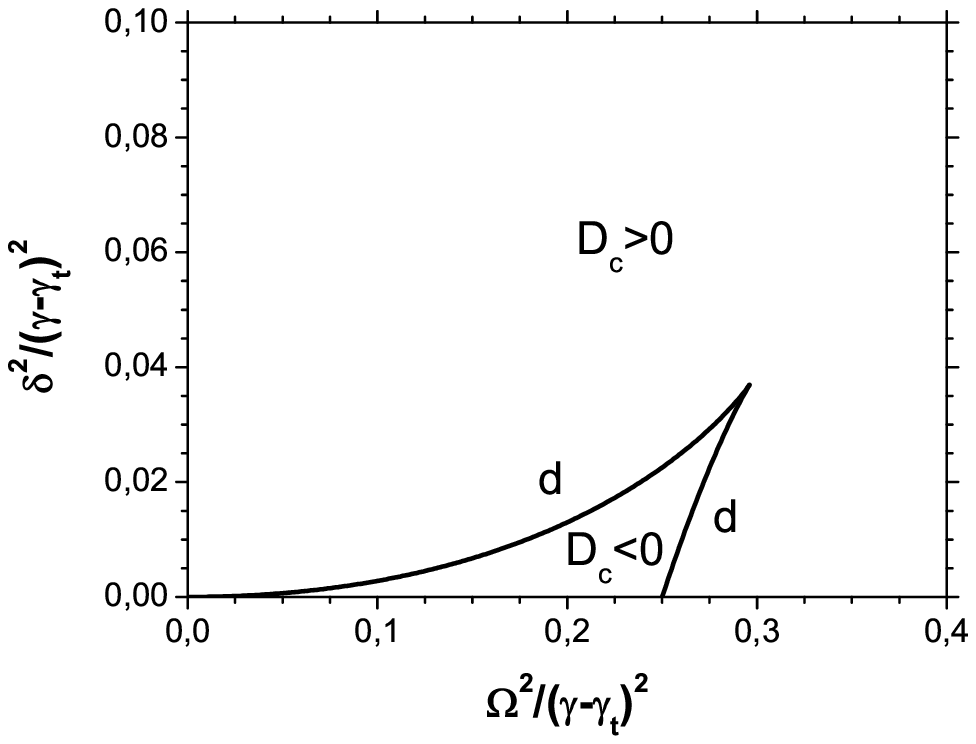,width=18cm,clip=0,angle=0}
\end{center}
\caption{}
\label{fgbnd}
\end{figure}

According to (\ref{pdc})
and figures \ref{fgdabz} and \ref{fgdab}  one has
$D_c(\alpha,\beta)>0$ for $\beta\geq 0$ and $\alpha> 1/27$. 
Therefore, the case of degenerate real roots is confined to
rather small detunings 
$|\delta|\lesssim 0,19245 |\gamma-\gamma_t|$ [see (\ref{alb})].
The dependence on $\beta$ in figures \ref{fgdabz} and \ref{fgdab} 
is extended to unphysical values
of $\beta<0$, which correspond to an imaginary magnitude of $\Omega$, 
in order to make the cubic dependence on $\beta$
[see the square bracket in (\ref{dscra})] transparent.
For $\alpha\in (0, 1/27)$, the zeros of $D_c(\alpha,\beta)$,
considered as a cubic function of $\beta\geq 0$, occur in {\em pairs}
(see also figure \ref{fgbnd}).
Indeed, the coefficients $a_j$ 
of a general cubic polynomial such as that in (\ref{dtp}) can
be alternatively expressed in terms of cubic roots 
$z_1$, $z_2$, and $z_3$ 
as $a_1=z_1z_2+z_1z_3+z_2z_3$ and 
$a_0=-z_1z_2z_3$. Given that the constant term of 
$h$ satisfies $a_0>0$ for $\alpha>0$, and combined with 
the existence of a negative real root of $h$, 
the real roots 
of $h$ have to come necessarily either with the signs $-++$ or $---$. 
The fact that in the case of $h$ one has $a_1<0$ for $\alpha<1/27$ 
then excludes the $---$ option.

According to figure \ref{fgbnd}, with {\em increasing} $\alpha$: 
\begin{itemize}

\item (i) the interval between the pair of zeros of  
$D_c(\alpha,\beta)$ along a $\beta$ trajectory {\em decreases} 
and 

\item (ii) 
the interval middle point shifts 
slightly to larger values of $\beta$ (see also figure \ref{fgdabz}). 

\end{itemize}
Since $\alpha<1/27$, and hence the condition (\ref{trc}) is not satisfied, 
each of the zeros of the pair corresponds to 
a {\em doubly} degenerate real root.
Along the boundary between the origin and the cusp point one has $h\equiv 0$
in (\ref{sdfa}), the term $\beta-2\alpha$ increases monotonically within
the boundaries
\begin{equation}
0\leq \beta-2\alpha\leq \frac{6}{27},
\end{equation}
and the doubly-degenerate root changes 
monotonically between $-\gamma_t$ and $-(\gamma+2\gamma_t)/3$. 
Along the boundary between the cusp point and $\beta=1/4$, 
the term $\beta-2\alpha$ in (\ref{sdfa}) continues to
increase monotonically within the interval
\begin{equation}
\frac{6}{27} \leq \beta-2\alpha \leq \frac{1}{4},
\end{equation}
and the doubly-degenerate root changes 
monotonically between $-(\gamma+2\gamma_t)/3$ and $-(\gamma+\gamma_t)/2$.

The case of a {\em triply} degenerate real root, which 
according to (\ref{trc}) occurs at $\alpha =1/27$ [see (\ref{pdc})], 
corresponds to the
case when the interval between the pairs of doubly degenerate 
zeros of $D_c$ along a given $\beta$-trajectory 
(parallel to the $x$-axis in figure \ref{fgbnd}) reduces to zero. 
The latter corresponds to the cusp point of the boundary 
shown in figure \ref{fgbnd}, which separates the $D_c>0$ and $D_c<0$ 
regions in the $(\Omega^2,\delta^2)$ plane.

\section{Steady state solutions and remaining numerical constants}
\label{sec:prp}
Cubic roots of $\Delta(p)$ determine the decay rates in (\ref{lplsf}).
The remaining part is to determine the numerical constants
in (\ref{lplsf}). Two of the numerical constants could be determined from 
the initial conditions for the Bloch variables
[see (\ref{lplsf})]
\begin{equation}
u_l(0) = 
\left\{
\begin{array}{ll}
A_0 + A_1  + A_2,  & D<0 
\\
A_0 + A_1  + A_2  + A_3,  & D > 0 
\\
A_0 + A_1  + A_2,  & D= 0 \mbox{ (double root)}
\\
A_0 + A_1,    & D=0 \mbox{ (triple root)}
\end{array} \right.
\label{lplsic}
\end{equation}
and for the first derivative of the Bloch variables
\begin{equation}
u_l'(0) = \left\{
\begin{array}{ll}
-\kappa_1 A_1  - b A_2  + A_3, & D<0 
\\
-\kappa_1 A_1  -\kappa_2 A_2 -\kappa_3 A_3,  & D > 0 
\\
-\kappa_1 A_1  -\kappa_2 A_2 + A_3,  & D= 0 \mbox{ (double root)}
\\
-\kappa_1 A_1  +A_2,  & D=0 \mbox{ (triple root)}.
\end{array} \right.
\label{lplsicd}
\end{equation}
In the latter case the left-hand side is provided by 
 (\ref{bleq}) taken at $t=0$.
A great deal of simplification can be achieved
when some of the constants (e.g. a 
{\em steady state solution} $A_0$ and $A_1$)
could be determined in advance, before one makes use
of the initial conditions.
Since $\Delta(p)$ has only nonzero roots, the steady state solution
$A_0$ can be determined from the PFD's 
listed by Eqs. (\ref{tdc}) as \cite{Tr}
\begin{equation}
A_0 = \frac{1}{\Delta(0)} \lim_{p\rightarrow 0} pf(p)
= \frac{1}{a_0}  \times \left\{
\begin{array}{cc} 
 \left(-\gamma \delta \Omega w_{eq}\right) , & l=1
\\
\gamma \Omega\,\gamma_t w_{eq} , & l=2
\\
 \gamma (\gamma_t^2+\delta^2) w_{eq},  & l=3
\end{array}\right.
\label{a0f}
\end{equation}
where $a_0$ is given by (\ref{a0}) and we have employed (\ref{fld})
in determining $\lim_{p\rightarrow 0} pf(p)$.
Provided that the root $x_1=-\kappa_1$ is nondegenerate (e.g. $D\ne 0$), 
the constant $A_1$ can be determined from the PFD's 
listed by Eqs. (\ref{tdc}) as \cite{Tr}
\begin{equation}
A_1 =  \lim_{p\rightarrow x_1} \frac{(p-x_1)f(p)}{\Delta(p)}
=  \frac{f(x_1)}{(x_1-z_2)(x_1-z_3)}\cdot
\label{a1f}
\end{equation}
Obviously, in the case $D> 0$ all the constants $A_j$, $j=1,2,3$
can be determined by a cyclic permutation of (\ref{a1f}).
For nearly degenerate roots one could instead use the expressions 
\begin{eqnarray}
A_2 &=& \frac{1}{x_3-x_2}\left[  x_3 (u_0-A_0-A_1) - (u_0'-x_1A_1)\right],\\
A_3 &=& \frac{1}{x_3-x_2}\left[ -x_2 (u_0-A_0-A_1) + (u_0'-x_1A_1)\right]. 
\label{a2a3}
\end{eqnarray}
In the case of a {\em triply} degenerate real root, one determines 
$A_1$ and $A_2$ straightforwardly from the initial conditions
(\ref{lplsic}) and (\ref{lplsicd}) and the knowledge of $A_0$
[see (\ref{a0f})]. $A_3$ could be determined from
\begin{equation}
A_3 =  \lim_{p\rightarrow -\kappa_1} \frac{(p+\kappa_1)^3 f(p)}{\Delta(p)}=f(-\kappa_1).
\label{a3f}
\end{equation}
One finds that the substitution of 
$x_{1,2,3}=-\kappa_1$ from   (\ref{tdr}), for $p$ in (\ref{fld}) 
amounts to replacing
\begin{equation}
(p+\gamma) \rightarrow \frac{2}{3}\, (\gamma-\gamma_t),\hspace*{1.5cm}
(p+\gamma_t) \rightarrow -\frac{1}{3}\, (\gamma-\gamma_t).
\end{equation}
Thus in any case it is possible to determine $A_0$ and one
of the constants $A_1$, $A_3$ directly from the knowledge of the
roots $0$ and $x_1=-\kappa_1$ of the product $p\Delta(p)$.

So far we have ignored a nearly trivial case of $\gamma_t=\delta=0$, 
in which case $a_0=0$ [see (\ref{a0})]. 
In general, one of the roots of $\Delta(p)$ 
is $\kappa_1=0$ if $a_0=0$ [see (\ref{dtp})].
Then $\Delta(p)$ in (\ref{dtp}) factorizes into a product of $p$ and a 
quadratic polynomial, and all the roots of $\Delta(p)$ 
can be straightforwardly obtained.
A necessary modification of the PFD, and of the recurrences 
for the coefficients $A_j$, 
to the case of a doubly degenerate real root $p=0$ is rather straightforward.
Eqs. (\ref{tdcz}) are modified to
\begin{equation}
\frac{f(p)}{\Delta(p)} =
\left\{
\begin{array}{ll}
 \frac{A_0}{p} + \frac{A_1}{p^2}+\frac{A_2(p+b) +A_3}{(p+b)^2 + s^2},
& D< 0
\\
 \frac{A_0}{p} + \frac{A_1}{p^2}+\frac{A_2}{p+\kappa_2}+\frac{A_3}{p+\kappa_3},
& D>0 
\\
 \frac{A_0}{p} + \frac{A_1}{p^2}+\frac{A_2}{p+\kappa_2}+\frac{A_3}{(p+\kappa_2)^2},
& D=0 \mbox{ (double root)}
\end{array}\right.
\label{tdcz}
\end{equation}
Numerical constants are determined as follows.
Instead of $A_0$ by Eq. (\ref{a0f}), one determines
\begin{equation}
A_1 = \frac{1}{\Delta(0)} \lim_{p \rightarrow 0} p^2f(p)
= \frac{1}{z_2z_3}  \times \left\{
\begin{array}{cc} 
 \left(-\gamma \delta \Omega w_{eq}\right) , & l=1
\\
\gamma \Omega\,\gamma_t w_{eq} , & l=2
\\
 \gamma [\gamma_t^2+\delta^2]w_{eq},  & l=3
\end{array}\right.
\label{a01f}
\end{equation}
Since it is no longer possible to obtain $A_0$
through  (\ref{a0f}), as the second constant 
for $D>0$ in (\ref{tdcz}) one could determine
\begin{equation}
A_3 =\frac{1}{2}\left(\frac{f(z)}{z} +\frac{f(\bar{z})}{\bar{z}}\right)
= \mbox{Re } \left(\frac{f(z)}{z}\right),
\end{equation}
where $z$ is one of the complex conjugate roots
[discussed earlier in connection with  (\ref{bsd})].
Eq. (\ref{a1f}) applies for $A_j$ only for $j=2,3$.
In the case of a {\em doubly} degenerate real root $x_2=x_3$,
\begin{equation}
A_3 =  \lim_{p\rightarrow x_2} \frac{(p-x_2)^2 f(p)}{\Delta(p)}=
\frac{f(x_2)}{x_2}\cdot
\label{a32f}
\end{equation}
We have excluded here the
unrealistic undamped case of $\gamma=\gamma_t=0$, which reduces
to the precession of a classical gyromagnetic moment 
in a magnetic field \cite{FVH,AE}, and which 
comprises the case $a_2=a_0=0$ leading to a triply 
degenerate root of $\Delta(p)$.

In the absence of {\em pure phase} relaxation processes 
(e.g. atoms in a dilute vapor cell; single 
molecule in solid hosts at superfluid helium 
temperatures \cite{TLB}; localised surface plasmons),
a pure dephasing rate is absent.
Then the above expressions simplify according to 
the substitutions $\gamma \rightarrow 2\gamma_t$ and
$(\gamma-\gamma_t)^2 \rightarrow \gamma_t^2$.

\section{Discussion}
\label{sec:disc}

\subsection{PFD over {\em real} number field}
Our treatment is unique in that it takes full advantage of the
PFD over {\em real} number field, which 
made it possible to find and classify all analytic solutions.
Surprisingly enough, the PFD has been nowhere mentioned in the context 
of Torrey's solution of the Bloch equations for a 
two-level system \cite{Tr,McH,AE,Ya,NJ} 
and has not been used in its full generality.
Surprisingly enough, earlier works \cite{Tr,HM} can be characterized by 
taking into account only one of the possible PFD's listed in 
Eqs. (\ref{tdc}) (see Eq. (41) of Ref. \cite{Tr}).

Laplace transform combined with  the PFD over {\em complex} number field 
has been employed to solve related problems of the
Bloch equations for a three-level system by Bernard et al \cite{BFT}
 [see Eq. (4) therein] and five-level kinetics by De Vries and  Wiersma 
[cf Eq. (A6) in Appendix A of Ref. \cite{VW}]. However, by making
use of the PFD in the {\em complex domain} one can no longer guarantee
that all the numerical constants in (\ref{pfd}) are {\em real} numbers.
The complex PFD obscures the fact that although any linear combination 
$U=\chi_1 e^{-\kappa_1 t} + \chi_2 e^{-\kappa_2 t}$
with real $\kappa_1,\, \kappa_2>0$ and $\chi_1,\, \chi_2\ne 0$ can be recast as 
$V=\xi_1 e^{-b t} \cosh (st) + \xi_2 e^{-b t} \sinh (st)$ with some real 
$b$ and $s$, the reverse is only possible for $b\pm s >0$.
However, if the latter condition holds, then $V$  becomes an awkward recasting
of $U$ which obscures that $V$ reduces to a sum of two
 simple exponential decays (cf Ref. \cite{NJ}).

\subsection{Earlier work}
Torrey \cite{Tr} did not additionally make use of Cardano's formula. 
His general solution was confined to the case $D<0$, when 
the characteristic determinant $\Delta(p)$ has a pair of complex conjugate
roots. Therefore Torrey \cite{Tr} missed the solutions corresponding 
to the parameter range $D\geq 0$. Hore and McLauchlan formally determined 
the cubic roots by Cardano's formula \cite{HM}. However, solution given by their 
Eq. (3) was limited to $D>0$. Indeed, coefficients $A_j$
given by their Eq. (4) become {\em singular} for degenerate roots. 
Furthermore, although the denominator in the expression for $A_1$
in their Eq. (4) is formally correct for $D>0$ [cf our (\ref{a1f})],
its numerator does not appear to be equal to the value of $f_l(-\kappa_1)$
defined by Eqs. (\ref{fld}).
Additionally, Hore and McLauchlan solution misses a constant term 
[see Eq. (3) in Ref. \cite{HM} with the second functional 
form of solutions of the Bloch equations in our (\ref{lplsf})].

Noh and Jhe \cite{NJ} have correctly pointed at the incompleteness
of Torrey's solution \cite{Tr}. They employed a slightly asymmetric form of 
the Bloch equations, which resulted when $u$ and $v$ in (\ref{uvwdf})
were defined without the prefactor of two.
However, they did not solve the Bloch equations {\em ab-initio} and
did not make use of the Laplace transform. 
Instead they employed a trial Ansatz
and implicitly employed Cardano's formula to classify different solutions.
Their solution is only limited
to the special (although the most important and the most studied) case of 
$w_0=w_{eq}=-1$.
In their approach, Noh and Jhe \cite{NJ} were required to look 
at the initial conditions
for the second derivative of the Bloch variables,
in spite that the Bloch equations (\ref{bleq}) do only contain the
first derivatives. Neither Noh and Jhe \cite{NJ} made a
connection between $D_c$ in Cardano's formula and the 
discriminant $D$ of a cubic polynomial [see (\ref{ddf}) and (\ref{vdmd})]. 
In the case of $D>0$, Noh and Jhe solution is expressed in terms 
of hyperbolic sine and cosine 
(see Eq. (12) in Ref. \cite{NJ}) - they overlooked 
that it can be simplified to a sum of 
three exponentially decreasing terms.
Additionally, they did not identify 
different parameter ranges of $D\lesseqqgtr 0$
in terms of the intrinsic parameters of the  Bloch equations
[see our (\ref{dscr}), (\ref{qlzc}), (\ref{ddr})].

\subsection{Transients and unorthodox solutions}
\begin{figure}[tbp]
\begin{center}
\epsfig{file=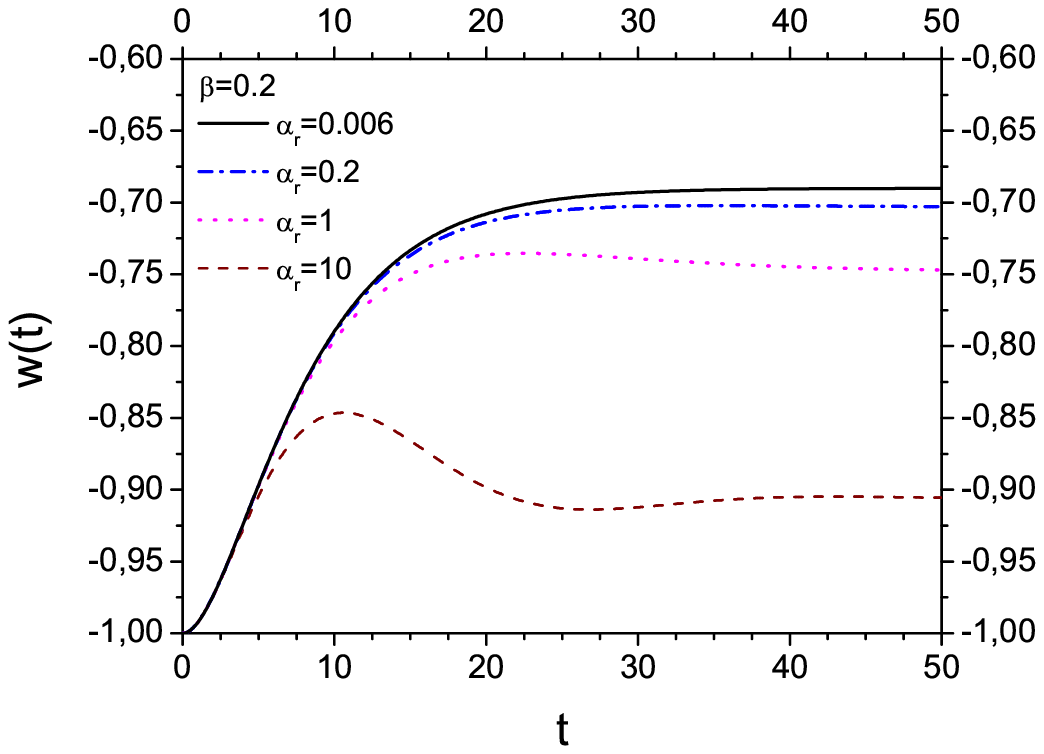,width=18cm,clip=0,angle=0}
\end{center}
\caption{}
\label{fgwdp}
\end{figure}

It has been shown that unorthodox solutions of the Bloch equations
are confined to rather small detunings 
$\delta^2\lesssim (\gamma-\gamma_t)^2/27$ and 
small field strengths $\Omega^2\lesssim 8 (\gamma-\gamma_t)^2/27$, which
are readily accessible experimentally.
Figure \ref{fgwdp} shows that, regarding time dependence
of the Bloch variables, there is a very smooth transition 
between the respective regions of $D_c<0$  and $D_c>0$. 
Even if  $\alpha_r$ were an order of magnitude larger than
 the boundary value of $\alpha_r \approx 0.35$ for  $\beta=0.2$, 
the resulting time-dependence of $w(t)$ would still resemble a 
featureless exponentially damped curve. The only indication
that one is outside the $D_c<0$ region is that the steady-state value 
of $w$ becomes marginally smaller than the curve
maximum. Therefore, in the parameter range of unorthodox solutions
and within a substantially larger parameter subrange of Torrey's solutions
proximal to the unorthodox solutions,  the Bloch variables
approach their respective steady states without
reaching any significantly higher values in a transient region.

\section{Conclusions}
\label{sec:conc}
A complete classification of unorthodox solutions of the Bloch equations
for a two-level system in time-harmonic driving classical field, 
which do not have the form of familiar 
Torrey's exponentially damped harmonic oscillations, has been provided.
Parameter range of the unorthodox solutions has been shown 
to be readily accessible experimentally.
Time dependence of unorthodox solutions is characterized by 
rather featureless exponentially damped behaviour.
The unorthodox solutions are essential for a reliable 
description of many different magnetic resonance \cite{Bl,Tr,FVH,HM,MK,JW,ACS} 
and quantum optics two-level systems \cite{FVH,McH,WLK,AE,Ya,BMO,BTL,TLB,TJB,BNB}. 
The complete set of solutions could also provide a testing ground
for general operator techniques involving
exponential solutions of differential 
equations for a linear operator \cite{Mgn,Fer,Anl,BCO}.
A F77 code used to generate plots here is freely available \cite{Cd}.

\newpage

\appendix

\section{Partial fraction decomposition over the field of real numbers}
\label{sc:pfe}
Suppose there exist 
{\em real} polynomials $f(x)$ and $g(x)\ne 0$, such that
\begin{equation}
    h(x) = \frac{f(x)}{g(x)}\cdot
\end{equation}
By removing the leading coefficient of $g(x)$, 
we may assume without loss of
generality that $g(x)$ is a polynomial whose 
leading coefficient is one
(i.e. monic polynomial). 
By the fundamental theorem of algebra, we can write
\begin{equation}
    g(x) = (x-x_1)^{j_1}\cdots(x-x_m)^{j_m}(x^2+b_1x+c_1)^{k_1}
                     \cdots(x^2+b_nx+c_n)^{k_n},
\label{fta}
\end{equation}
where $x_1,\ldots, x_m, b_1,\ldots, b_n, c_1,\ldots, c_n$ 
are all {\em real} numbers with $b_i^2 - 4c_i < 0$, and 
$j_1,\ldots, j_m$, $k_1,\ldots, k_n$ are positive integers. 
The $x_j$'s correspond to {\em real} 
roots of $g(x)$, and the terms 
$(x^2 + b_ix + c_i)$ [the so-called {\em irreducible quadratic 
factors} of $g(x)$] correspond to pairs of complex conjugate roots 
of $g(x)$.
The partial fraction decomposition of $h(x)$ is
\begin{equation}
    h(x) = \frac{f(x)}{g(x)} = q(x) + 
    \sum_{i=1}^m\sum_{r=1}^{j_i} \frac{A_{ir}}{(x-x_i)^r} 
       + \sum_{i=1}^n\sum_{r=1}^{k_i} \frac{B_{ir}x+C_{ir}}{(x^2+b_ix+c_i)^r}\cdot
\label{pfd}
\end{equation}
Here $q(x)$ is a (possibly zero) polynomial, and 
the $A_{ir}$, $B_{ir}$, $C_{ir}$, $b_i$, and $c_i$ 
are all {\em real} constants.
A further information on the partial fraction decomposition
can be found in Ref. \cite{Pfdw}.

\section{Summary of inverse Laplace transform formulae}
\label{ap:iltf}
\begin{eqnarray}
\mathcal{L}^{-1} \{s^{-1}\} &=& \Theta(t),
\nonumber\\
\mathcal{L}^{-1} \left\{\frac{1}{s+\alpha}\right\} 
                     &=& \Theta(t) e^{-\alpha t},
\nonumber\\
\mathcal{L}^{-1} \left\{\frac{\omega}{(s+\alpha)^2 +\omega^2}\right\} 
               &=& \Theta(t) e^{-\alpha t}\sin(\omega t),
\nonumber\\
\mathcal{L}^{-1} \left\{\frac{s+\alpha}{(s+\alpha)^2 +\omega^2}\right\} 
           &=& \Theta(t) e^{-\alpha t}\cos(\omega t),
\nonumber\\
\mathcal{L}^{-1} \left\{\frac{1}{(s +\alpha)^n}\right\} 
           &=& \Theta(t) \frac{t^{n-1}}{(n-1)!}\, e^{-\alpha t},
\nonumber\\
\mathcal{L}^{-1} \{s^{-2}\} &=& \Theta(t)\, t,
\end{eqnarray}
where $\Theta(t)$ is the Heaviside step function
whose value is zero for negative argument and 
one for positive argument.


\newpage

\begin{center}
{\large\bf Figure captions}
\end{center}

\vspace*{1.4cm}

\noindent {\bf Figure 1. -}
Plots of $D_c(\alpha,\beta)$ as a function of $\beta$ for various values
of the detuning $\delta^2$ in the units of $(\gamma-\gamma_t)^2/27$. 
One has $D_c(\alpha,\beta)> 0$ for $\beta\geq 0$,
provided that $\alpha_r>1$ ($\alpha> 1/27\approx 0.\overline{037}$).

\vspace*{0.4cm}

\noindent {\bf Figure 2. -}
Zoom-out view of Figure 1 showing cubic curves
of the plots of $D_c(\alpha,\beta)$ as a function of $\beta$.

\vspace*{0.4cm}

\noindent {\bf Figure 3. -}
The boundary separating $D_c>0$ and $D_c<0$ regions in the
$(\Omega^2,\delta^2)$-plane. The cusp point corresponds to 
the triply-degenerate root.
Along the boundary between the origin and the cusp point, the 
doubly-degenerate root changes 
monotonically between $-\gamma_t$ and $-(\gamma+2\gamma_t)/3$, whereas
between  the cusp point and $\beta=1/4$  the doubly-degenerate root changes 
monotonically between $-(\gamma+2\gamma_t)/3$ and $-(\gamma+\gamma_t)/2$,
respectively.

\vspace*{0.4cm}

\noindent {\bf Figure 4. -}
Time dependence of of the Bloch variable $w(t)$ for $\gamma=0.4$, $\gamma_t=0.1$,
and $\beta=0.2$ for various values
of the detuning $\delta^2$ in the units of 
$(\gamma-\gamma_t)^2/27$. The first two smallest values 
of $\alpha_r$ correspond to the $D_c<0$ region
($\beta=0.2$ corresponds the boundary value of $\alpha\approx 0.01299$, 
or $\alpha_r \approx 0.35$)
and the remaining two values of $\alpha_r$ correspond to the Torrey ($D_c>0$) 
region. Note in passing a very smooth transition in the time dependence
of $w(t)$ between the respective
regions of $D_c<0$  and $D_c>0$.

\newpage

\newpage

\newpage

\newpage

\begin{thebibliography}{99}



\bibitem{Bl}
F. Bloch,
Nuclear induction,
Phys. Rev. 70 (1946) 460-474.


\bibitem{Tr}
H.C. Torrey,
Transient nutations in nuclear magnetic resonance,
Phys. Rev. 76 (1949) 1059-1068.


\bibitem{FVH}
R.P. Feynman, F.L. Vernon, Jr., R.W. Hellwarth, 
Geometrical representation of the Schr\"{o}dinger equation 
for solving maser problems,
J. Appl. Phys. 28 (1957) 49-52.


\bibitem{HM}
P.J. Hore, K.A. McLauchlan,
CIDEP and spin relaxation measurements by flash photolysis EPR methods,
J. Magn. Reson. 36 (1979) 129-134. 


\bibitem{MK}
P.K. Madhu, A. Kumar,
Direct Cartesian-space solutions of 
generalized Bloch equations in the rotating frame,
J. Magn. Reson. A 114 (1995) 201-211.


\bibitem{McH}
S.L. McCall, E.L. Hahn,
Self-induced transparency,
Phys. Rev. 183 (1969) 457-485.


\bibitem{WLK}
J.H.-S. Wang, J.M. Levy, S.G. Kukolich, J.I. Steinfeld,
Microwave transient nutation measurements of relaxation 
in OCS and NH${}_3$,
Chem. Phys. 1 (1973) 141-148.


\bibitem{AE}
L. Allen, J.H. Eberly, Optical Resonance and Two-level
Atoms, John Wiley \& Sons, New York, 1975.


\bibitem{Ya}
A. Yariv, Quantum Electronics, John Wiley \& Sons, New York, 1975.


\bibitem{BMO}
T. Basch\'{e}, W.E. Moerner, M. Orrit, H. Talon,
Photon antibunching in the fluorescence of a single 
dye molecule trapped in a solid,
Phys. Rev. Lett. 69 (1992) 1516-1519.


\bibitem{BTL}
C. Brunel, P. Tamarat, B. Lounis, J. Plantard, M. Orrit,
Driving the Bloch vector of a single molecule: 
Towards a triggered single photon source,
Comptes Rendus de l'Academie de Sciences 
- Serie IIb: Mecanique, Physique, Chimie, 
Astronomie 326 (1998) 911-918.


\bibitem{TLB}
P. Tamarat, B. Lounis, J. Bernard, M. Orrit,
S. Kummer, R. Kettner, S. Mais, T. Basch\'{e},
Pump-probe experiments with a single molecule: 
ac-Stark effect and nonlinear optical response,
Phys. Rev. Lett. 75 (1995) 1514-1517.


\bibitem{TJB}
P. Tamarat, F. Jelezko, C. Brunel, A. Maali, B. Lounis, M. Orrit,
Non-linear optical response of single molecules,
Chem. Phys. 245 (1999) 121-132.


\bibitem{BNB}
B. Butscher, J. Nipper, J.B. Balewski, L. Kukota, V. Bendkowsky,	
R. L\"{o}w, T. Pfau,
Atom-molecule coherence for ultralong-range Rydberg dimers,
Nature Physics 6 (2010) 970-974.


\bibitem{FMR}
E.B. Flagg, A. Muller, J.W. Robertson, S. Founta, 
D.G. Deppe, M. Xiao, W. Ma, G.J. Salamo, C.K. Shih,
Resonantly driven coherent oscillations in a solid-state quantum emitter,
Nature Physics 5 (2009) 203-207.


\bibitem{DS}
S.K. Dutta {\em et al},
Multilevel effects in the Rabi oscillations of a Josephson phase qubit,
Phys. Rev. B 78 (2008) 104510.


\bibitem{JW}
F. Jelezko, J. Wrachtrup,
Single defect centres in diamond: A review,
phys. stat. sol. (a) 203 (13) (2006) 3207-3225.


\bibitem{ACS}
I. Aharonovich, S. Castelletto, D.A. Simpson, C.-H. Su, A.D. Greentree, 
and S. Prawer,
Diamond-based single-photon emitters,
Rep. Prog. Phys. 74 (2011) 076501.


\bibitem{TCC}
J.M. Taylor, P. Cappellaro, L. Childress, L. Jiang, D. Budker, 
P.R. Hemmer, A. Yacoby, R. Walsworth, M.D. Lukin,
High-sensitivity diamond magnetometer with nanoscale
resolution,
Nature Physics 4 (2008) 810-816.

\bibitem{ML}
J.R. Maze {\em et al},
Nanoscale magnetic sensing with an individual electronic spin in diamond,
Nature 455 (2008) 644-647.


\bibitem{BJW}
G. Balasubramanian {\em et al},
Nanoscale imaging magnetometry with diamond spins under ambient conditions,
Nature 455 (2008) 648-651.



\bibitem{Hr}
J.P. Hornak, The Basics of NMR.
Available online
\verb|http://www.cis.rit.edu/htbooks/nmr/inside.htm|.


\bibitem{Pfdw}
A further information on the partial fraction decomposition
can be found on 
\verb|http://en.wikipedia.org/wiki/Partial_fraction|.


\bibitem{Cf}
A nice exposition of the Cardano's formula and the 
roots of a cubic polynomial can be found at
\verb|http://mathworld.wolfram.com/CubicFormula.html| and
\verb|http://en.wikipedia.org/wiki/Cubic_polynomial|.


\bibitem{Cr}
A. Ekert, Girolamo Cardano - the gambling scholar,
Phys. World, May 2009, 36-40.


\bibitem{AMr}
A. Moroz, On a fully compensated formula for cubic roots (unpublished).


\bibitem{Dsc}
See for instance
\verb|http://en.wikipedia.org/wiki/Discriminant|.


\bibitem{NJ}
H.-R. Noh, W. Jhe,
Analytic solutions of the optical Bloch equations,
Opt. Commun. 283 (2010) 2353-2355.


\bibitem{Bt}
B. de Bartolo, Optical Interactions in Solids
John Wiley \& Sons, New York, 1968, pp. 431-442.


\bibitem{BFT}
J. Bernard, L. Fleury, H. Talon, M. Orrit, 
Photon bunching in the fluorescence from single molecules: 
A probe for intersystem crossing,
J. Chem. Phys. 98 (1993) 850-860.


\bibitem{VW}
H. de Vries, D.A. Wiersma,
Fluorescence transient and optical free induction 
decay spectroscopy of pentacene in mixed crystals 
at $2$ K, 
J. Chem. Phys. 70 (1979) 5807-5823.


\bibitem{Mgn}
W. Magnus, 
On the exponential solution of differential 
equations for a linear operator,
Commun. Pure Appl. Math. 7 (1954) 649-673.


\bibitem{Fer}
F. Fer, 
R\'{e}solution de l'equation matricielle $\dot{U} = pU$ 
par produit infini d'exponentielles matricielles,
Bull. Classe Sci. Acad. Roy. Bel. 44 (1958) 818-829.



\bibitem{Anl}
P. Aniello, 
A new perturbative expansion of the time evolution 
operator associated with a quantum system,
J. Opt. B: Quantum Semiclass. Opt. 7 (2005) S507-S522.


\bibitem{BCO}
S. Blanes, F. Casas, J. A. Oteo, J. Ros, 
Magnus and Fer expansions for matrix differential 
equations: the convergence problem,
J. Phys. A 31 (1998) 259-268.

\bibitem{Cd}
A F77 code used to generate plots here is freely available at
\verb|http://www.wave-scattering.com/bloch.html|


\end{thebibliography}
\end{document}